\documentclass[twocolumn,aps,pre,showpacs,floatfix,superscriptaddress]{revtex4-1}
\usepackage{graphics}
\usepackage{graphicx}
\usepackage{amsmath}
\usepackage{amssymb}
\usepackage{ulem}
\usepackage{color}

\begin{document}
\title{Spacing distribution functions for the one-dimensional point-island model with irreversible attachment}

\author{Diego Luis Gonz\'alez}
\email[]{dgonzal2@umd.edu}
\affiliation{Department of Physics, University of Maryland, College Park, Maryland 20742-4111  USA}
\author{Alberto Pimpinelli}
\email[]{alpimpin@univ-bpclermont.fr}
\altaffiliation[On leave from: ]{LASMEA, UMR 6602 CNRS/Universit\'e Blaise Pascal -- Clermont 2,
F-63177 Aubi\`ere cedex, France}
\affiliation{Department of Physics, University of Maryland, College Park, Maryland 20742-4111  USA}
\affiliation{French Embassy, Consulate General of France, Houston, Texas 77056 USA}
\author{T. L. Einstein}
\email[]{einstein@umd.edu}
\affiliation{Department of Physics, University of Maryland, College Park, Maryland 20742-4111  USA}

\date{\today}

\begin{abstract}
We study the configurational structure of the point-island model for epitaxial growth in one dimension. In particular, we calculate the island gap and capture zone distributions. Our model is based on an approximate description of nucleation inside the gaps. Nucleation is described by the joint probability density $p^{XY}_{n}(x,y)$, which represents the probability density to have nucleation at position $x$ within a gap of size $y$. Our proposed functional form for $p^{XY}_{n}(x,y)$ describes excellently the statistical behavior of the system. We compare our analytical model with extensive numerical simulations. Our model retains the most relevant physical properties of the system.

\end{abstract}


\maketitle

\section{Introduction}
The study of the theory of the submonolayer film growth is of great general interest, in part because its application in the construction of microelectronic devices. In epitaxial growth, monomers are deposited onto a substrate at a constant rate. The monomers diffuse until they are captured by an island or another diffusive monomer. The islands are just clusters of immobile monomers. The size of an island depends on the number of monomers which have attached to it. Thus, the island size increases in time due to the capture of diffusing monomers. These kinds of systems exhibit many interesting non-equilibrium phenomena, as discussed in, e.g., Refs.~\cite{blackman,amar}

There are basically two kinds of models which have been developed to study these systems: In the extended island model, the islands occupy more than one site on the lattice, and their shape is not trivial \cite{amar,mulheran,amar1,ratsch,amar2,blackman2}. In the simpler point-island model, the islands just occupied one site in the lattice, and their size is simply the number of monomers which belong to the island  \cite{amar3,evans,amar4,amar5,amar6,amar7,ratsch1,evansgama,tokar}. In this paper, we focus on the point-island model, which is very accurate for low coverages. A major advantage of this simplification is that it produces better statistics than the extended island model.

In our model only the monomers are mobile, the islands formed by two or more monomers are completely static (stable). In the literature this condition is usually denoted ``critical nucleus size" $i=1$ (where $i$ is the size of the largest unstable island). In our irreversible-growth model, monomers must land or hop onto an already occupied site in order to be incorporated \cite{amar3,amar7}; in an alternative model features incorporation of monomers arriving at sites adjacent to the island \cite{evansgama}.

The quantities that are commonly used to describe the evolution of the point-island model are the density of monomers, $N_1$; the density of islands with size $j$, $N_j$; the rate of deposition of monomers, $F$; and the diffusion rate of monomers $D$. The evolution of this system is frequently described in terms of the coverage $\theta=F\,t$, where $t$ is the time. We restrict our studies to the aggregation regime, where there is a quasi steady-state.

We are particularly interested in the spacing (gap) distribution functions between islands $\hat{p}^{(k)}(S)$, the capture zone distribution $\hat{P}(S)$, and the island-island pair correlation function $G(r)$. As usual, $\hat{p}^{(k)}(S)dS$ is the probability that for an island at the origin we find another island at a distance between $S$ and $S+dS$, with the condition that there are $k$ additional islands inside the gap between them. The standard definition for the scaled spacing is $s=S/\left\langle S\right\rangle$, with $\left\langle S\right\rangle$ the average of $S$.
The scaled spacing distributions are given by
\begin{equation}\label{pns}
p^{(k)}(s)=\left\langle S\right\rangle \hat{p}^{(k)}(s \left\langle S\right\rangle).
\end{equation}
On the other hand, we use the definition for $\hat{P}(S)$ given in \cite{amar3}. Then, in our one-dimensional (1D) system, the capture zone of an island is simply the distance between the midpoints of the gaps to the left and to the right of the island.

The functional forms of $p^{(0)}(s)$ and $P(s)$ for arbitrary dimension and critical nucleus size have been the subject of recent discussion and some controversy~\cite{amar3,pimpinelli1,evans1,pimpinelli}.
A particular issue is whether the generalized Wigner surmise (GWS)

\begin{equation}\label{gws}
\hspace{-2mm} P_\beta(s)\! =\! a_\beta s^\beta {\rm e}^{-b_\beta s^2}\! , \; b_\beta\! =\!
\left[\frac{\Gamma({\beta+2\over2})}{\Gamma({\beta+1\over2})}\right]^2\! , \; a_\beta\! =\! \frac{2
b_\beta^{(\beta+1)/2}}{\Gamma({\beta+1\over2})},\hspace{-2mm}
\end{equation}

\noindent adequately describes the distribution and the reliability the simple relationship between $\beta$ and $i$ deduced with mean field \cite{pimpinelli1} and later refined with more sophisticated arguments \cite{pimpinelli}.   There is still no general consensus about them even in the case of $i=1$ in one dimension (1D).  In this paper we calculate analytically and numerically the spacing distribution functions $p^{(k)}(s)$ for $k\geq0$, a meaningful concept for capture zones in 1D but less so in higher dimensions.  All previous studies focus just in $p^{(0)}(s)$ \cite{blackman,amar4,pimpinelli1}. We specially focus in their functional forms in the limit of large and small values of $s$, where deviations from the form of Eq.~(\ref{gws}) were observed in painstaking computations in 1D and 2D, see Ref.~\cite{amar3}. We propose an analytical model to find approximate expressions for these functions. Our model is based on a detailed description of the nucleation mechanism, following principally Blackman and Mulheran~\cite{blackman}, hereafter BM. We also calculate the pair correlation function $G(r)$. In all cases our model is compared with several numerical simulations. This paper is organized as follows: In Sec.~2 we discuss briefly some implications of the 1D model. In Sec.~3 we provide an accurate description of the nucleation process. In Sec.~4 and Sec.~5, we calculate approximately the spacing distribution functions, while in Sec.~6 we calculate the island-island pair correlation function. \textcolor[rgb]{0.00,0.00,0.00}{In Sec.~7, we assess the viability of applying Eq.~(\ref{gws}) to experimental data.}  In Sec.~8, we give conclusions.

\section{1D Point Island Model}
In the 1D case, we have a ring divided into independent sections called gaps. Each gap starts and ends with an island. Within the gaps, there may be several monomers performing random walks. A monomer inside of a particular gap must eventually either merge with one of the islands at the ends of the gap or combine with another monomer to nucleate a new island. In no case can the monomer reach a different gap.

The spatial distribution of the islands, i.e., the distribution of the sizes of the gaps, is given by the spacing distribution functions $p^{(k)}(s)$. The simplest case corresponds to the nearest-neighbor distribution $p^{(0)}(s)$, which represents the probability density to find a gap with an scaled size $s$ between two islands with no additional island in between.

The pair correlation function $G(r)$ is related to $p^{(k)}(s)$ by
\begin{equation}\label{grp}
G(r)=\sum^{\infty}_{k=0}p^{(k)}(r).
\end{equation}

\begin{figure}[htp]
\begin{center}
\includegraphics[scale=0.35]{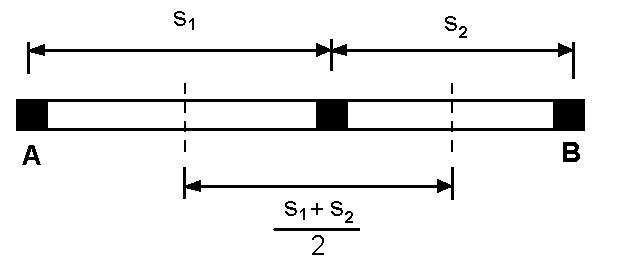}
\end{center}
\caption{Relation between $P(s)$ and $p^{(1)}$(s). Black blocks are islands, while white regions are gaps.}
\label{cpzp1}
\end{figure}

As noted, the length of the capture zone of an island is simply the distance between the midpoints of the gaps to the left and to the right of the island. In our 1D model, the next-nearest neighbor distribution $p^{(1)}(s)$ is related to $P(s)$ as follows. Consider Fig~\ref{cpzp1}, where the black squares represent islands. Let  $S_1+S_2$ be the distance between the islands $A$ and $B$. Then $\hat{p}^{(1)}(S_1+S_2)$ is the probability density to find a gap with size $S_1+S_2$ given that there is an additional island inside the gap. From its definition, it is clear that $p^{(1)}(s_1+s_2)$ is related to the capture zone distribution $P((s_1+s_2)/2)$ according to

\begin{equation}\label{cpzgo}
P(s)=\int^{\infty}_{0}dx\,p^{(1)}(x)\,\delta\left(s-\frac{x}{2}\right)=2\,p^{(1)}(2s),
\end{equation}
where $x=s_1+s_2$.

Thus, the capture zone distribution and the next-nearest spacing distributions are equivalent in one dimension.

\section{Spatial description of the nucleation}

\begin{figure*}[htp]
\begin{center}
$\begin{array}{cc}
\includegraphics[scale=0.3]{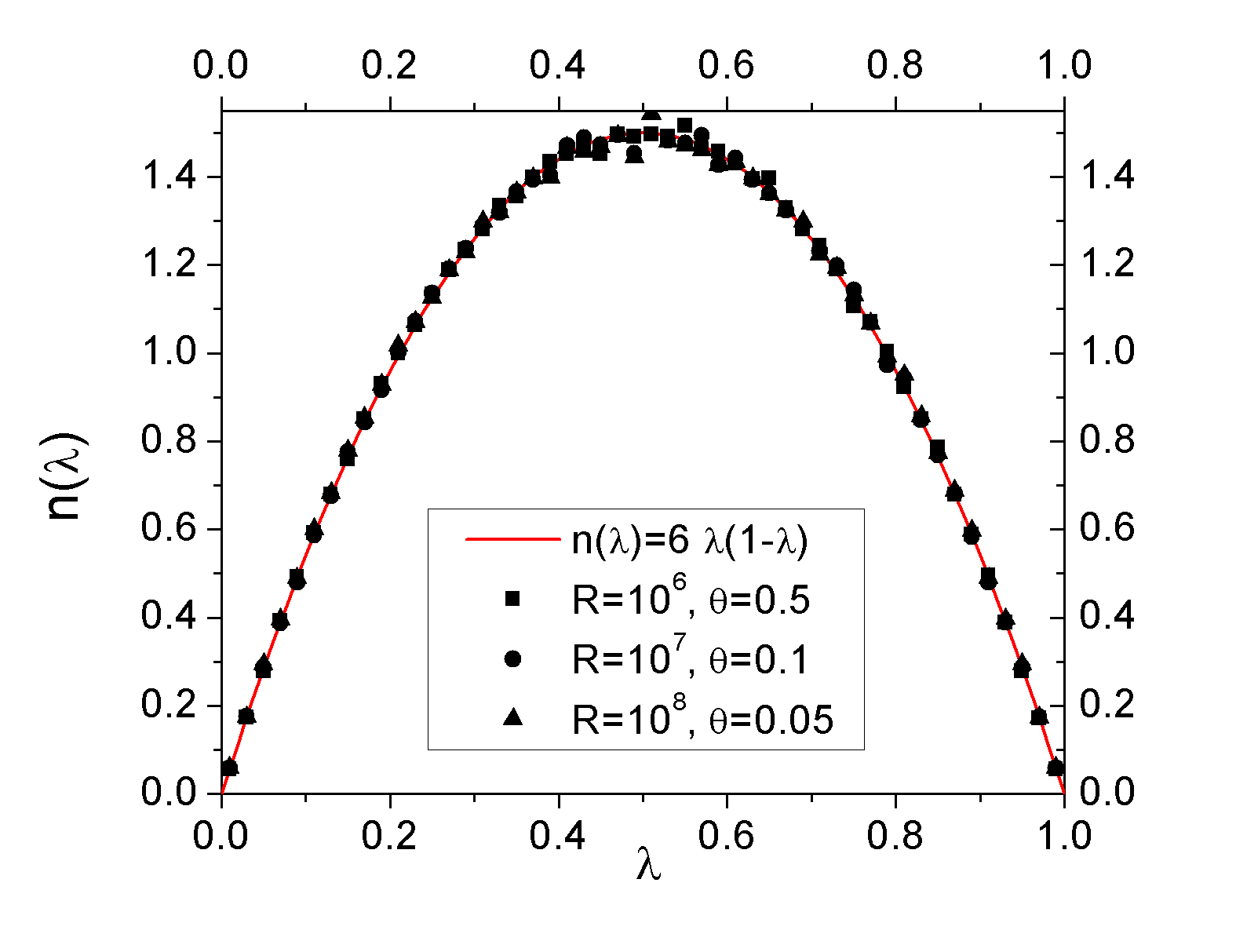}&
\includegraphics[scale=0.3]{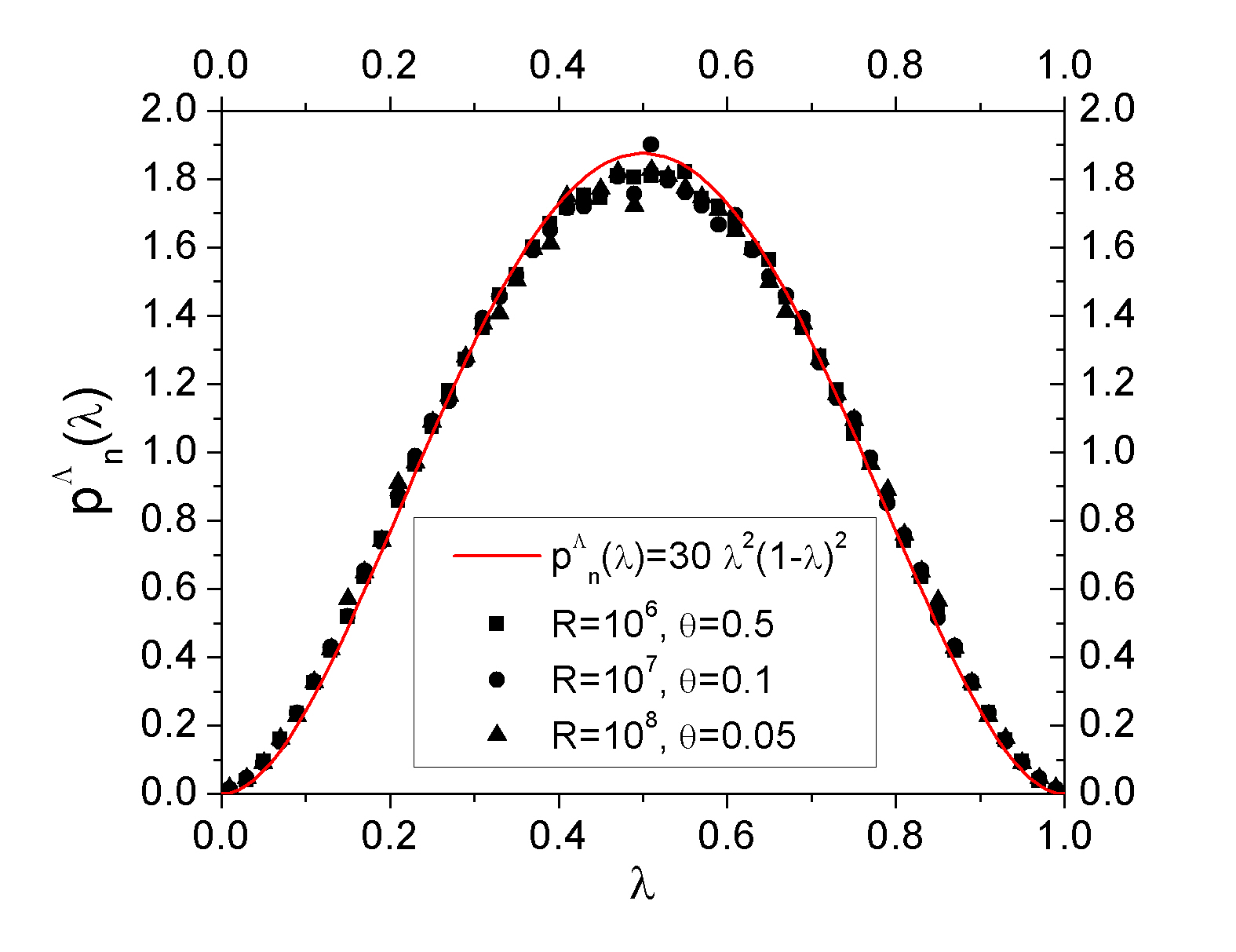}\\
(a) & (b) \\
\end{array}$
\end{center}
\caption{(Color online) Behavior of (a) the normalized reduced density $n(\lambda=x/y)$ (see Eq.~(\ref{nlambda})) and (b) the probability $p^{\Lambda}(\lambda=x/y))$ that a nucleation occurs at position $x$ in a gap of length $y$, for different values of $R$. The agreement between the analytical model and the numerical data is excellent overall. There are some slight differences near the maximum of $p^{\Lambda}(\lambda)$.}
\label{graph1}
\end{figure*}

In order to find an expression for $p^{(0)}(s)$, we must describe the creation mechanism of new gaps. To reach an appropriate description of the nucleation, we make the following definitions: Let $p^{XY}_{n}(x,y)$ be the joint probability density that a given new nucleation occurs at position $x$ inside a gap of length $y$. Of course $x<y$; otherwise $p^{XY}_{n}(x,y)=0$. Let $p^{X}_{n}(x)$ be the probability density that a given nucleation occurs at position $x$ inside a gap of any size. Then
\begin{equation}\label{pxd}
p^{X}_{n}(x)=\int^{\infty}_{x}dy\,p^{XY}_{n}(x,y).
\end{equation}
Similarly, the probability density $p^{Y}_{n}(y)$ that a given nucleation occurs anywhere inside a gap of length $y$ is given by
\begin{equation}\label{pyd}
p^{Y}_{n}(y)=\int^{y}_{0}dx\,p^{XY}_{n}(x,y).
\end{equation}
Finally, the probability density that, a nucleation occurs at position $x$ in a gap of length $y$ under the condition $\lambda=x/y$ is
\begin{eqnarray}\label{pld}
p^{\Lambda}_{n}(\lambda)&=&\int^{\infty}_{0}dy\int^{y}_{0}dx\,p^{XY}_{n}(x,y)\,\delta\left(\lambda-\frac{x}{y}\right)\nonumber\\
&=&\int^{\infty}_{0}dy\,y\,p^{XY}_{n}(\lambda y,y).
\end{eqnarray}

The formation of a new island depends directly on the probability that two monomers reach the same site of the lattice at the same time. This quantity is of course related to the density of monomers inside the gap. In BM, the average density of monomers, $n_1(x,y)$, inside of a gap of length $y$ was approximated by using its expression in the stationary state. In this regime there is quasi-equilibrium between the deposition of new monomers and their merging with existing islands. We neglect the deposition onto the top of islands. Then $n_1(x,y)$ satisfies a one-dimensional diffusion equation with the boundary conditions $n_1(0,y)=n_1(y,y)=0$. This leads to
\begin{equation}
n_1(x,y)=(2R)^{-1}x(y-x),
\end{equation}
with $R=D/F$~\cite{blackman}. By using this expression and the assumption that the probability of a new nucleation at $x$ is proportional to $n_1(x,y)^2$, we have
\begin{equation}
p^{\Lambda}_{n}(\lambda)=30\,\lambda^2(1-\lambda)^2,
\label{pld2}
\end{equation}

\noindent where $\lambda=x/y$ \cite{blackman}.  While we have neglected the interaction among monomers within a gap, a more refined estimate of $p^{\Lambda}_n(\lambda)$ can be made by taking them into account and using the full multiparticle density \cite{castellano,vardavas}.  We tested numerically the validity of this approximation. As seen in Fig.~\ref{graph1}, there is excellent agreement between the numerical data and the analytical equations. Note that in Fig.~\ref{graph1} we plotted the normalized reduced density $n(\lambda)$ given by
\begin{equation}\label{nlambda}
n(\lambda)=\int^{\infty}_{0}dy\int^{y}_{0}dx\,\frac{12\,R\,n_1(x,y)}{\mu_3}\,p^{(0)}(y)\,\delta\left(\lambda-\frac{x}{y}\right)
\end{equation}
instead of $n_1(x,y)$. Here $\mu_3$ is the $3^{rd}$ moment of $p^{(0)}(y)$ and the factor $12\,R/\mu_3$ guarantees the correct normalization of $n(\lambda)$ in the interval $0\leq\lambda\leq1$. From its definition, it is clear that $n(\lambda)$ gives the average density of monomers inside of a gap at the relative position $\lambda=x/y$. A major advantage of $n(\lambda)$ is that it does not depends explicitly on the length of the gap, in contrast to $n_1(x,y)$. The numerical simulation of the point-island model was carried out following the standard procedures~\cite{blackman,evans,amar4}. In our simulations we took a lattice with length $L = 2\times10^5$ sites (with periodic boundary conditions)
and gathered statistics from over $5000$ realizations.

Since $p^{\Lambda}_{n}(\lambda)$ vanishes for $\lambda=0$ and 1, a new island is unlikely to form near the boundaries of a gap. This is the origin of the effective repulsive force between adjacent islands. As we shall shortly see, this implies that for small values of $s$, $p^{(0)}(s)\propto s^{\alpha}$. We expect that $\alpha=2$ because $p^{\Lambda}(\lambda)\propto \lambda^2$ for $\lambda \rightarrow 0$.

\begin{figure}[htp]
\begin{center}
\includegraphics[scale=0.3]{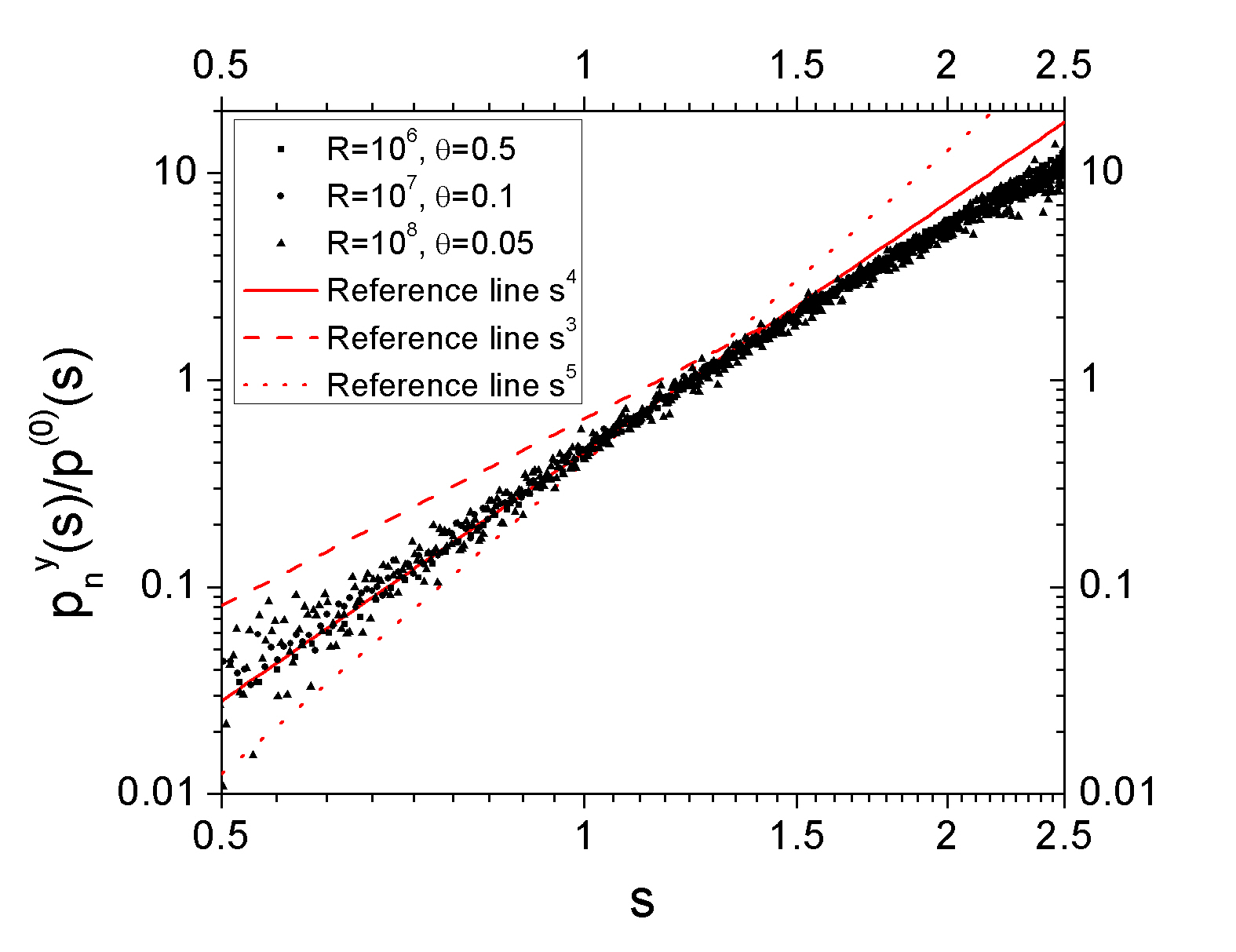}
\end{center}
\caption{(Color online) Behavior of the quotient $p^{Y}_{n}(s)/p^{(0)}(s)$. It seems that there are two different regimes, one for small gaps and the other for large gaps.}
\label{pysmall}
\end{figure}

BM also proposed that $p^{Y}_{n}(y)/F_M(y)\propto y^5$, where $F_M(y)$ is the number of gaps with size $y$ given that there are $M$ gaps. $F_M(S)$ is naturally related to $p^{(0)}(s)$ by $F_M(S)=(M^2/L)p^{(0)}(s)$. We calculate numerically this quotient for different values of $R$. As shown in Fig.~\ref{pysmall}, our numerical results suggest that $p^{Y}_{n}(y)/F_M(y)\propto y^{\gamma}$ with $\gamma\approx3$ for $s>1.7$ and $\gamma\approx4$ for $s<1.7$ rather than $\gamma=5$. A similar result was found in Ref.~\cite{evansgama} for submonolayer deposition in 2D. Their numerical results suggests that $\gamma(s)\approx (4+s)/(2+s)$. Hence, for small gaps we have $\gamma\approx2$ while for large gaps $\gamma\approx1$.

Implicitly in BM, the probability density $p^{Y}_{n}(y)$ was written as
\begin{equation}\label{pxyblack}
p^{Y}_{n}(y)=\left(\int^{y}_{0}dx\,n_1^2(x,y)\right) p^{(0)}(y)\propto y^5p^{(0)}(s).
\end{equation}
Underlying Eq.~(\ref{pxyblack}) are the following two approximations: First, the integral is based on the law of mass action; Fig.~\ref{graph1} justifies this. Second, BM supposed that  $p^{Y}_{n}(y)$ can be written as the product of the probability to have a nucleation inside a gap of size $y$ and the number of gaps $F_M(y)$ with this size. Our numerical results, shown in Fig.~\ref{pysmall}, do not support that simplification.

Following our previous observations, $p_n^{XY}(x,y)$ can be written as the independent product of the probability $p^Y_n(y)$ to nucleate inside a gap with size $y$ and the conditional probability $p^{\Lambda}_n(x/y)/y$ that this nucleation occurs at $x$. After some algebra we find

\begin{equation}\label{epnxy}
p^{XY}_{n}(x,y)=\frac{30}{\mu_{\gamma}}x^2(y-x)^2 y^{\gamma-5}p^{(0)}(y),
\end{equation}
where $\mu_{\gamma}$ is the $\gamma^{th}$ moment of $p^{(0)}(y)$. It is easy to show that Eq.~(\ref{epnxy}) satisfies the form of $p^{\Lambda}_n(\lambda)$ given in BM. The condition $p^{Y}_{n}(y)/p^{(0)}(y)\propto y^{\gamma}$ is satisfied as well. From Eqs.~(\ref{pxd}), (\ref{pyd}) and (\ref{epnxy}), it is straightforward to show
\begin{equation}\label{pn}
p_n^Y(s)=\frac{s^{\gamma}}{\mu_{\gamma}}p^{(0)}(s)
\end{equation}
and
\begin{equation}\label{pn1}
p^X_n(s)=\frac{30\,s^2}{\mu_{\gamma}}\int^{\infty}_{s}\,\frac{dy}{y^{5-\gamma}} p^{(0)}(y)(y-s)^2.
\end{equation}
Equations (\ref{pn}) and (\ref{pn1}) describe the mechanism of formation of new islands.  Equation (\ref{pn}) shows that the probability of nucleation inside a gap of size $y$ is the product of the number of such gaps and $y^{\gamma}$. As we will see $\gamma\neq5$ in contradiction to Eq.~(\ref{pxyblack}). Equation (\ref{pn1}) shows how $p^X_n(s)$ depends on the number of monomers inside the gap; for small $s$, $p^X_n(s)\propto s^2$, consistent with the law of mass action.  The vanishing of $(y-s)^2$ as $y \rightarrow s$ and $s^2$ as $s \rightarrow 0$ reflects that nucleation does not occur at the gap borders because ipso facto the monomer concentration vanishes there.

\begin{figure*}[t!]
\begin{center}
$\begin{array}{cc}
\includegraphics[scale=0.3]{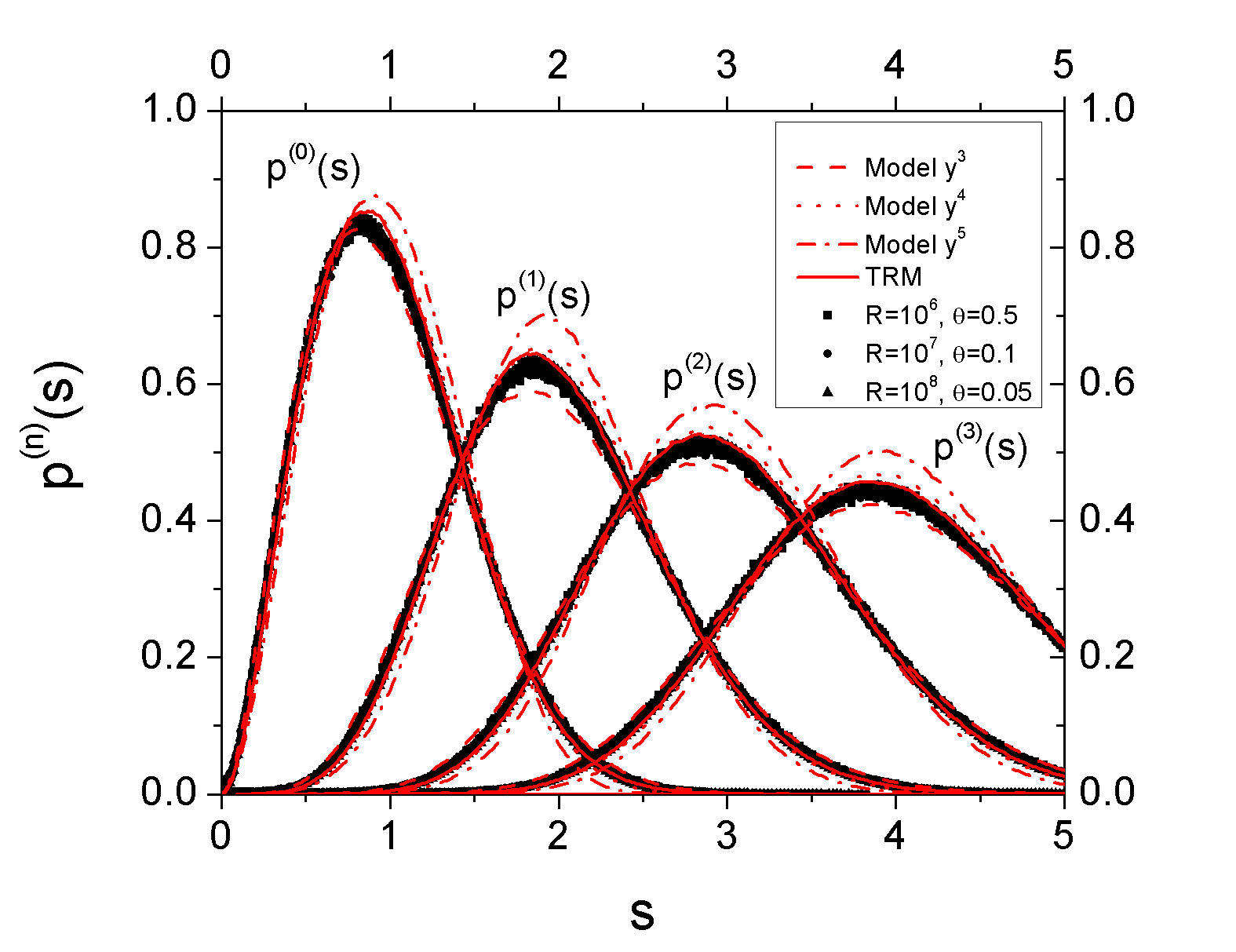}&
\includegraphics[scale=0.3]{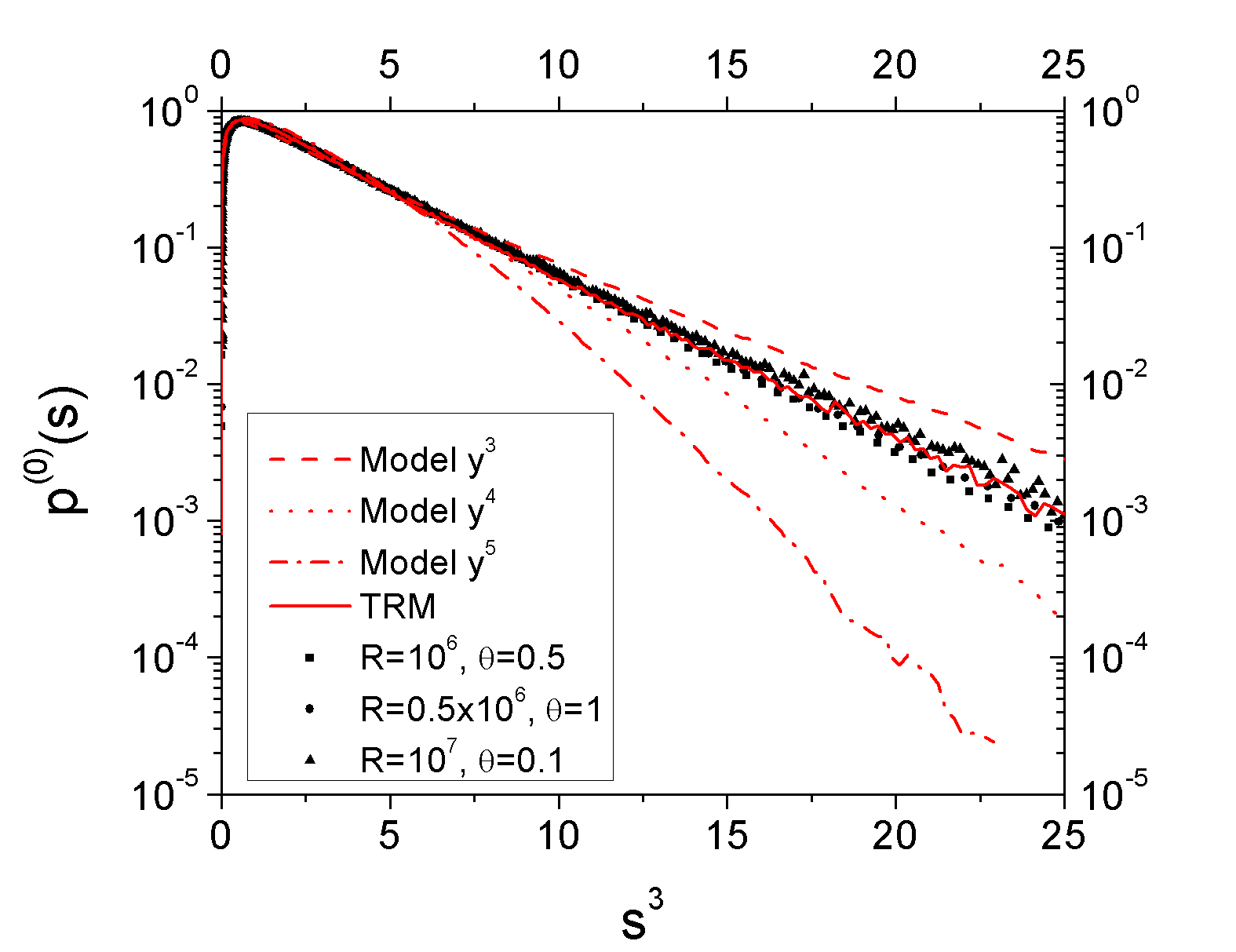}\\
(a) & (b) \\
\end{array}$
\end{center}
\caption{(Color online) Comparison of the statistical solution of Eq.~(\ref{fint}) and the numerical simulation of the point-island model with $i=1$. In (a) we show the first four spacing distribution functions and (b) shows the behavior of $p^{(0)}(s)$ for large values of $s$.}
\label{p01D}
\end{figure*}

\begin{figure*}[t!]
\begin{center}
$\begin{array}{cc}
\includegraphics[scale=0.3]{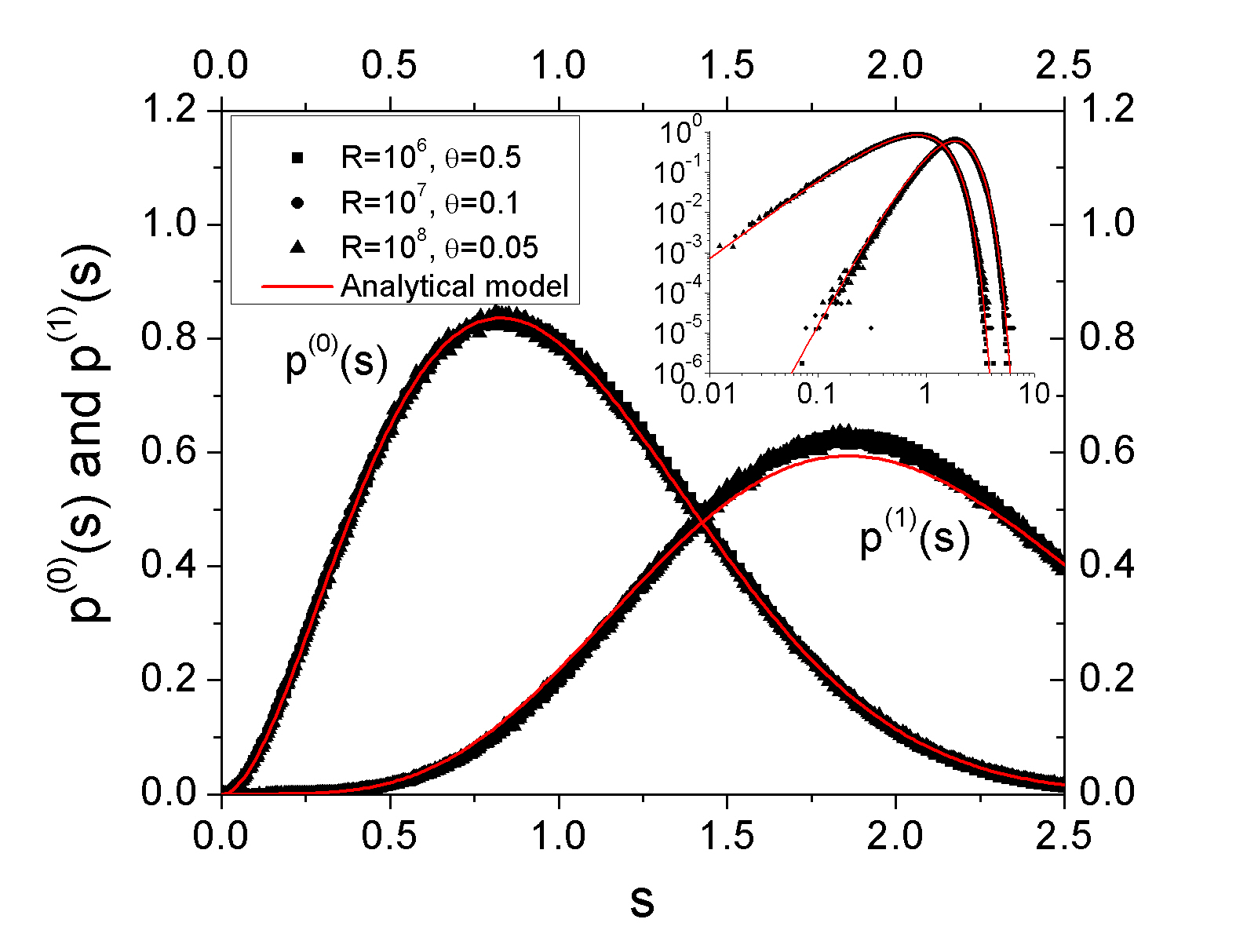}&
\includegraphics[scale=0.3]{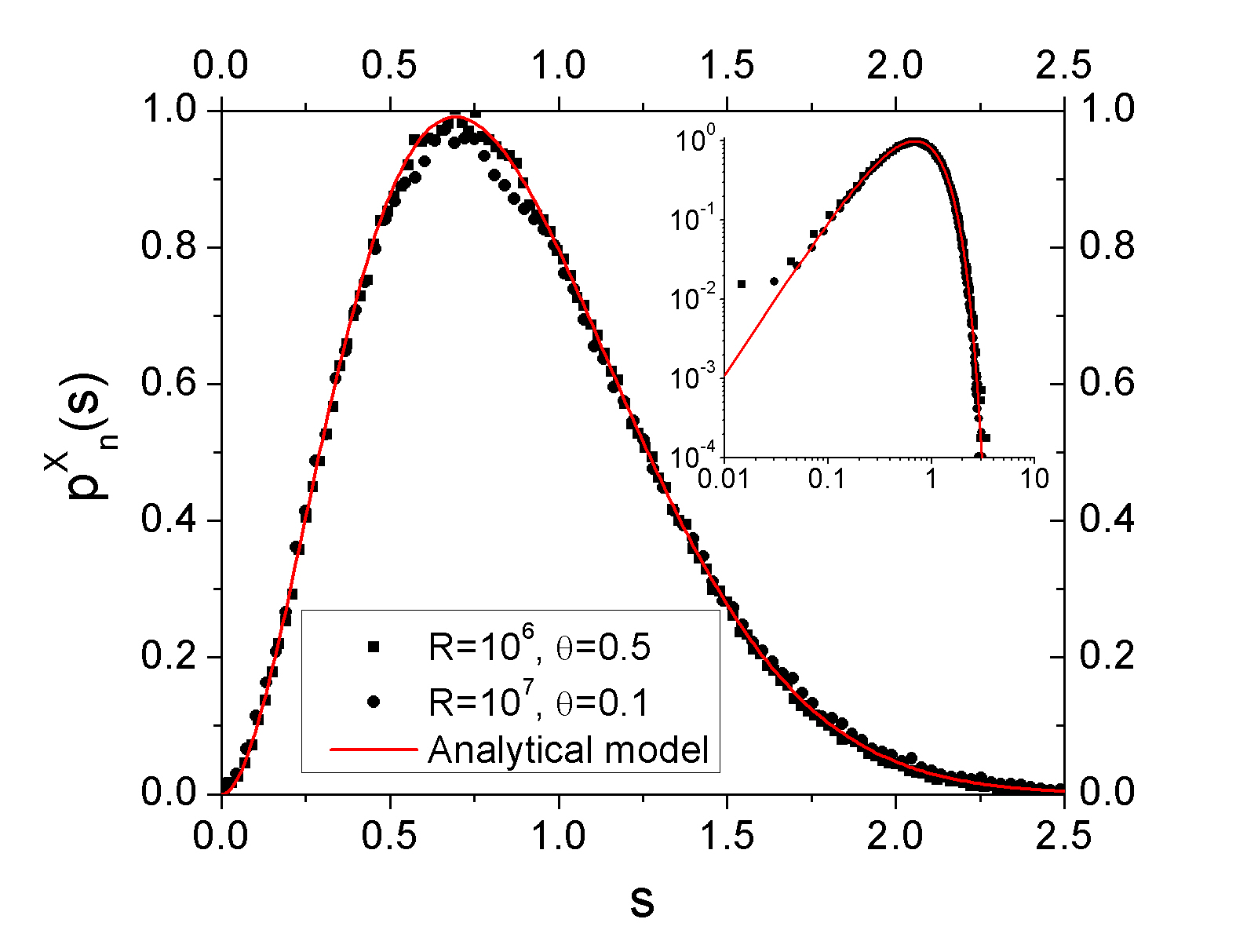}\\
(a) & (b) \\
\end{array}$
\end{center}
\caption{(Color online) The behavior of $p^{(k)}(s)$ and  $p^{X}_n(s)$ are shown in (a) and (b), respectively. There is excellent agreement between Eqs.~(\ref{paprox}) and (\ref{px}), respectively, and the numerical data from the simulation of the point-island model.}
\label{ps}
\end{figure*}

\section{Nearest-neighbor spacing distribution $p^{(0)}(s)$}

Following BM, it is possible to find an analytical expression which relates $p^{(0)}(y)$ with $p_n^Y(s)$ and $p^X_n(s)$: The effect of one single nucleation is described by
\begin{equation}
F_{M+1}(S)-F_M(S)=-p^Y_{n}(S)+R_n(S),
\end{equation}
with
\begin{eqnarray}
R_n(S)&=&\int^{\infty}_{x}dy\int^{Y}_{0}dx\,p^{XY}_{n}(x,y)\left(\delta(x-S)+\delta(x-y+S)\right)\nonumber\\
&=&\int^{\infty}_{x}dy\,\left(p^{XY}_{n}(S,y)+p^{XY}_{n}(S-y,y)\right).
\end{eqnarray}
Due to the symmetry of $p^{\Lambda}(\lambda)$ around its maximum, $p^{XY}_{n}(S,y)=p^{XY}_{n}(S-y,y)$.  Thus,
\begin{equation}
F_{M+1}(S)-F_M(S)=-p^Y_{n}(S)+2\,p^{X}_{n}(S).
\end{equation}
Finally, taking into account the relation between $F_M(S)$ and $p^{(0)}(s)$, we find
\begin{equation}\label{ed1d}
s \frac{dp^{(0)}(s)}{ds}+2\,p^{(0)}(s)=-p^Y_{n}(s)+2\,p^{X}_{n}(s).
\end{equation}
From Eqs.~(\ref{pn}), (\ref{pn1}) and (\ref{ed1d}) we can write
\begin{equation}\label{difp}
s\,\frac{dp^{(0)}(s)}{ds}+\left(2+\frac{s^{\gamma}}{\mu_{\gamma}}\right)p^{(0)}(s)=\frac{60\,s^2}{\mu_{\gamma}}\int^{\infty}_{s}dy\frac{(y-s)^2}{y^{5-\gamma}} p^{(0)}(y).
\end{equation}
This integro-differential equation can be written in the following integral form
\begin{equation}\label{fint}
p^{(0)}(s)=\frac{2\,e^{-\frac{s^{\gamma}}{\gamma \mu_{\gamma}}}}{s^2}\int^{s}_{0}
dy\,y\,p^{X}_{n}(y)\,e^{\frac{y^{\gamma}}{\gamma \mu_{\gamma}}},
\end{equation}
where $p^{X}_{n}(y)$ is given by Eq.~(\ref{pn1}). BM attributed the exponential tail of $p^{(0)}(s)$ to the fact that $p^Y_{n}(s)\propto y^{\gamma}$, regardless of the form of $p^{\Lambda}_n(\lambda)$. We recall that, in their case, $\gamma=5$. Consequently, they claim that, for large values of $s$, $p^{(0)}(s)\propto \exp(-s^{5}/5\,\mu_{\gamma})$.  Furthermore, the exponential tail of $p^{(0)}(y)$ depends on the value of $\gamma$ in the ratio $p^{Y}_n(s)/p^{(0)}(s)\propto s^{\gamma}$ regardless the form of $p_n^{\Lambda}(\lambda)$. This means that the fragmentation process for large values of $s$ depends on the probability of choosing the gap to fragment, while for small values of $s$, it depends exclusively on the probability of choosing the place of fragmentation.

From Eqs.~(\ref{pn1}) and (\ref{fint}) it can be shown that in the limit $s\ll1$, $p^{(0)}(s)=(15/\mu_{\gamma})s^2+O(s^3)$ and $p^X_n(s)=(30/\mu_\gamma)s^2+O(s^3)$. In the opposite limit $s\rightarrow\infty$ we have $p^{(0)}(s)\propto s^{-2}\exp(-s^{\gamma}/\gamma\,\mu_{\gamma})$ (see Appendix).

As noted in BM, an integro-differential equation like Eq.~(\ref{difp}) is hard to solve analytically or numerically. However, it can be solved easily in an statistical way: A line of length $L$ is taken and an array of points on the line is generated. Each new point is introduced via weighted random number generation. First, a pair of existing points is selected between which to introduce the new point. The selection is weighted by the $\gamma^{th}$ power of the separation of the points. We used $\gamma=3,4$ and 5. Then, the actual position in the gap for insertion $x$ is selected with weighting given by the function $p^{\Lambda}_{n}(x/y)$.  We use $L=10000$ and gather data for $100$ particles and $50000$ realizations. Henceforth, we call this procedure the statistical solution of Eq.~(\ref{difp}).

We also seek a simple approximate expression for $p^{(0)}(s)$ which can be used instead of the statistical solution of the integro-differential equation. In order to find it, we focus on the known moments of $p^{(0)}(s)$. The first two moments are chosen to satisfy the standard normalization conditions. In Ref.~\cite{blackman}, it was shown that in the aggregation regime the density of monomers, $N_1$, is related to the islands density, $N$, according to
\begin{equation}\label{q3}
\mu_3=12\,N^2N_1\,R \approx 1.6,
\end{equation}
where $\mu_3$ is the third moment of $p^{(0)}(s)$. There is another condition which can be extracted from $p^{(0)}(s)$. In the aggregation regime, the spacing distribution functions have only a weak dependence on time. Because of this we can interpret the point-island model as a one-dimensional system in which the particles (islands) interact with their nearest neighbors under a potential $v(s)$. In the limit of small coverage and small values of $s$, we can write $p^{(0)}(s)\approx \mathrm{\rm exp}[-v(s)/k_{B}\,T]$. For simplicity, we henceforth set the temperature to be $k_{B}\,T=1$. It follows that $v(s)\approx-2\,\mathrm{ln}(s)$. It is easy to check the validity of this approximate equation from our results for $p^{(0)}(s)$. The average interaction energy per particle $u$ is constant; using the maximum entropy method (MEM) \cite{jaynes,wu,kumar}, we can approximate $u$ by
\begin{equation}
u=\frac{1}{\left\langle S\right\rangle}\int^{\infty}_{0}dS\,v(S)\,p_{\rm MEM}^{(0)}\left(\frac{S}{\left\langle S\right\rangle}\right).
\end{equation}
\noindent To determine $p_{\rm MEM}^{(0)}(s)$, we take note of four independent conditions: two normalization conditions, the third moment of $p^{(0)}(s)$ and the average of $v(s)$. Then
$p_{\rm MEM}^{(0)}(s)$ is given by
\begin{equation}
p_{\rm MEM}^{(0)}(s)=e^{C_1+C_2s+C_3s^3+C_4\mathrm{ln}\left(s^2\right)},
\end{equation}
where $C_i$ are constants. We conclude that an approximate analytical form for $p^{(0)}(s)$ is
\begin{equation}\label{paprox}
p_{\rm MEM}^{(0)}(s)= A\,s^2e^{-B\,s-C\,s^3},
\end{equation}
with $A$, $B$ and $C$ constants to be determined.

By using Eqs.~(\ref{ed1d}) and (\ref{paprox}), we can calculate $p^X_n(s)$, finding  straightforwardly
\begin{equation}\label{px}
p^X_n(s)\approx \frac{p^{(0)}(s)}{2 \mu_{\gamma}}\left(s^{\gamma} +4\,\mu_{\gamma} - \mu_{\gamma} B\,s -3\,\mu_{\gamma} C\,s^3\right).
\end{equation}

Fig.~\ref{p01D}(a) shows the spacing distribution functions $p^{(k)}(s)$ calculated from the simulation of the point-island model and the results of the statistical solution of Eq.~(\ref{difp}) with $\gamma=3,4$ and 5. In all cases the agreement is good for small values of $s$ and still good for intermediate values of $s$ with $\gamma=3$ and 4. However, none of them describe correctly $p^{(0)}(s)$ for large values of $s$ (see Fig.~\ref{p01D}(b)). To improve the model we choose $\gamma=4$ for $s<1.7$ and $\gamma=3$ for $s>1.7$, as prompted by Fig.~\ref{pysmall}. This two-regime model, hereafter termed TRM, gives an excellent fit with the point-island model even for large values of $s$ (see Fig.~\ref{p01D}).

Fig.~\ref{p01D}(b) reveals the important result that $p^{(0)}(s)$ does not decay as a Gaussian but instead more like $\exp(-C\,s^3)$. This form differs from the ones proposed previously in Refs.~\cite{blackman,pimpinelli1} but is consistent with Eq.~(\ref{paprox}).

In Fig.~\ref{ps} we see that Eq.~(\ref{paprox}) gives excellent results for $p^{(0)}(s)$ even for large and small values of $s$. In Fig.~\ref{ps}(a) we show the results of a least-squares fit to calculate the extra parameter of Eq.~(\ref{paprox}). In Fig.~\ref{ps}(b) we use Eqs.~(\ref{px}) to calculate $p^X_n(s)$. The agreement with the numerical data is excellent.

\section{Higher spacing distribution functions}

Now we turn to higher-spacing distribution functions. Equation (\ref{cpzgo}) and the weak correlation between gaps sizes yields the standard approximate equation for $P(s)$
\begin{equation}\label{cz1D}
P(s)\approx2\int^{2s}_{0}dx\,p^{(0)}(x)p^{(0)}(2s-x),
\end{equation}
This equation was used satisfactorily in Refs.~\cite{blackman,pimpinelli1} to calculate $P(s)$ from $p^{(0)}(s)$. Numerical evidence there supported the validity of this approximation, which is usually called the independent interval approximation (IIA). The solutions given by Eqs.~(\ref{paprox}) and (\ref{px}) will be called the analytical model. In Ref.~\cite{gonzalez4} there is additional numerical evidence confirming the validity of the IIA in this kind of model. In IIA the expression analogous to Eq.~(\ref{ed1d}) for $p^{(1)}(s)$ is

\begin{equation}\label{edp21d}
s\frac{dp^{(1)}(s)}{ds}+3\,p^{(1)}(s)=-q^Y_{n}(s)+4\,q^{X}_{n}(s),
\end{equation}
where we use
\begin{equation}\label{qy}
q^Y_{n}(s)=\int^{s}_{0}dx\,p^{(0)}(s-x)p^Y_{n}(s)
\end{equation}
and
\begin{equation}\label{qx}
q^{X}_{n}(s)=\int^{s}_{0}dx\,p^{(0)}(s-x)p^{X}_{n}(s).
\end{equation}
Equation (\ref{qy}) represents the probability density that a new nucleation event occurs inside a gap (such as the one shown in Fig.~\ref{cpzp1}), while Eq.~(\ref{qx}) is the analog of $p^{X}_{n}(s)$ for this kind of gap. In the IIA  $q^Y_{n}(s)$ and $q^{X}_{n}(s)$ are written as convolution products. The $p^{(1)}(s)$ shown in Fig.~\ref{ps} was calculated by using Eq.~(\ref{paprox}) and the IIA. We find excellent agreement even for small and large values of $s$. There are some small discrepancies (less than 6\%) near the maximum of $p^{(1)}(s)$.

It can be shown that in IIA, the spacing distribution for small values of $s$ can be written as  $p^{(k)}(s)\propto s^{\alpha_k}$ with $\alpha_k=3\,n+2$. In particular, this means that $P(s)\propto s^{5}$. This result agrees with BM's results but differs from the exponent reported in Ref.~\cite{pimpinelli1}.

\section{Pair correlation functions}
\begin{figure}[t]
\begin{center}
\includegraphics[scale=0.35]{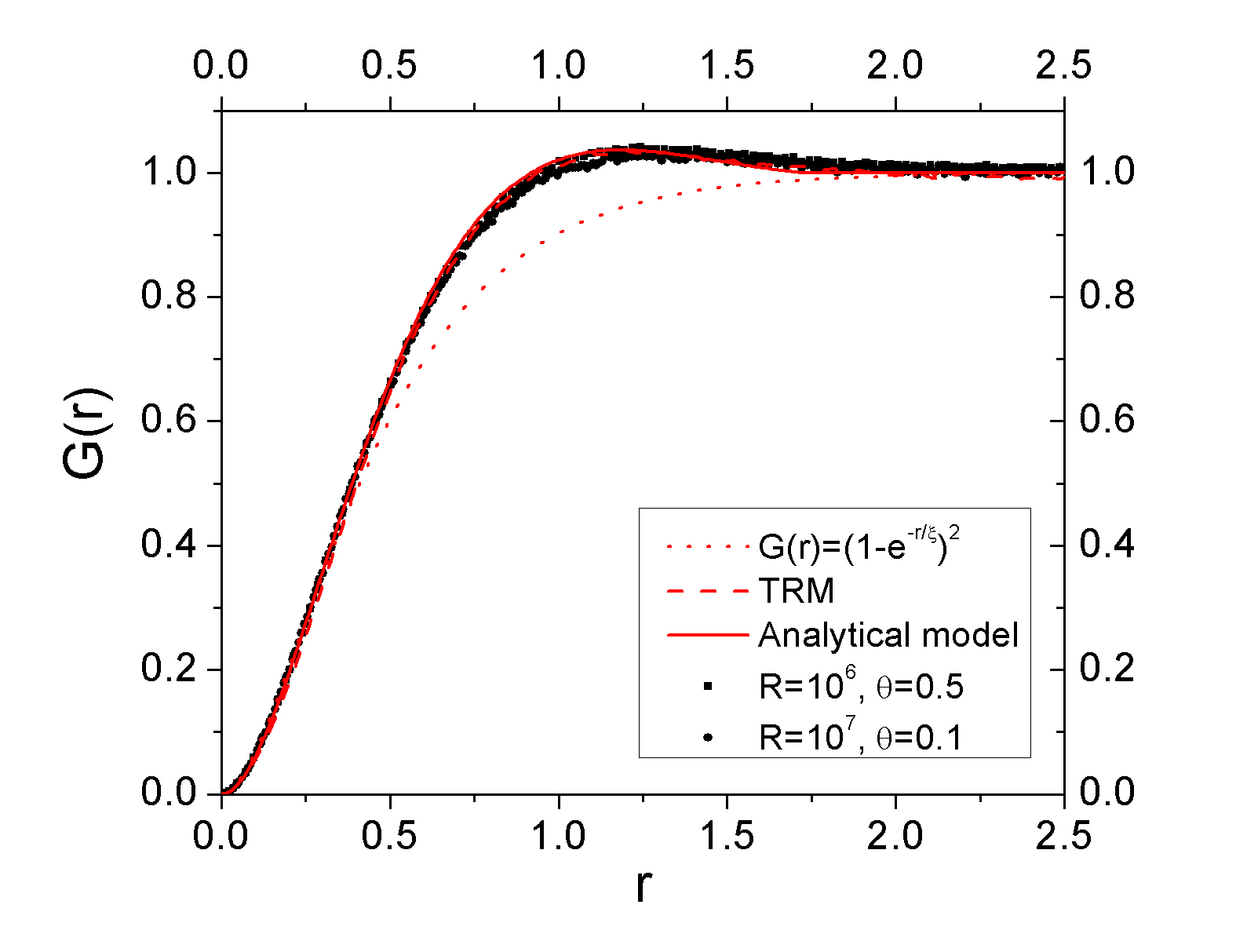}
\end{center}
\caption{(Color online) Pair correlation function  $G(r)$.}
\label{grf}
\end{figure}

The island-monomer pair correlation function $g(r)$ is given by
\begin{equation}
g(r)=\frac{N_1(r)}{N_1},
\end{equation}
where $N_1(r)$ describes the monomer density at a distance $r$ from an specified island. From this definition of $g(r)$ it is clear that $g(r)\propto r$ for small values of $r$ and $g(r)\rightarrow1$ for $r \rightarrow \infty$. In Refs.~\cite{blackman,bales},  $N_1(r)$ was calculated by using a diffusion equation corresponding to modeling the sea of islands as an uniform sink. In this way, they found
\begin{equation}\label{gre}
g(r)=1-e^{-r/\xi},
\end{equation}
with $\xi$ the average distance a monomer travels before being captured by an island or another monomer. The island-island pair correlation function is defined as $G(r)=N(r)/N$, with $N(r)$ the density of islands at a distance $r$ from the center of a specific island. In this case we have $G(r)\propto\,r^2$ for small values of $r$; $G(r)$ naturally satisfies the same condition as $g(r)$ for large values of $r$. It is well known \cite{evansg,evansr} that, at least in two-dimensional systems, $G(r)$ and $g(r)$ satisfy the approximate relation
\begin{equation}\label{gG}
G(r)\approx g^{i+1}(r).
\end{equation}
If we extrapolate this observation to our 1D case, we have $G(r)\approx g^2(r)$. By using Eqs.~(\ref{gre}) and (\ref{gG}) we can obtain an expression for $G(r)$. These equations predict that $g(r)$ and $G(r)$ grow monotonically. However, our numerical results, displayed in  Fig.~\ref{grf}, show a weak oscillation near $r=1$. This oscillation has been observed experimentally in two-dimensional systems \cite{evansr,bressler,bartel}. Both models---the one given by the statistical solution of Eq.~(\ref{difp}) with our TRM and the one given by Eq.~(\ref{paprox}) plus the IIA---reproduce correctly this oscillation. The behavior of $G(r)$ is shown in Fig.~\ref{grf}. In the analytical model we use Eqs.~(\ref{grp}) and (\ref{paprox}). The solution for $G(r)$ obtained from the statistical solution of Eq.~(\ref{difp}) is plotted as well. The agreement between these calculations and the numerical simulation is evident. The calculations even reproduce the weak oscillations.

In a test of the validity of Eq.~(\ref{gG}), measuring $g(r)$ directly from the simulation, we did not find good agreement with the numerical results.  Since $G(r)\approx g^2(r)$ arises from the assumption that $dN(r)/dt\propto N_1(r)^{i+1}$, apparently the law of mass action is not reliable for the whole ring, even though it is valid for the gaps.

\section{Viability of Generalized Wigner Surmise (GWS) for Experimental Data}

\subsection{Comments from preceding analysis}

\begin{figure*}[t!]
\begin{center}
$\begin{array}{cc}
\includegraphics[scale=0.3]{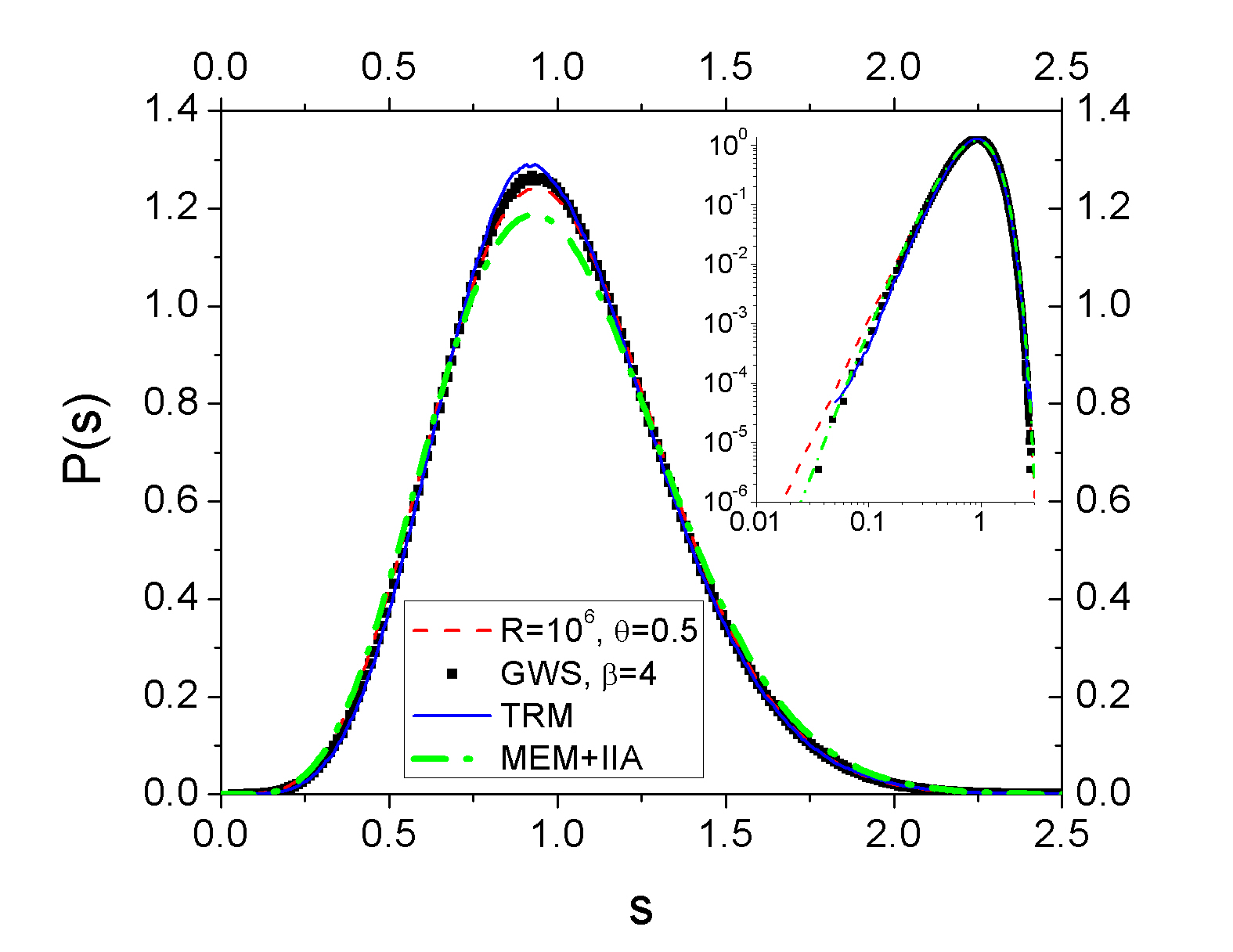}&
\includegraphics[scale=0.3]{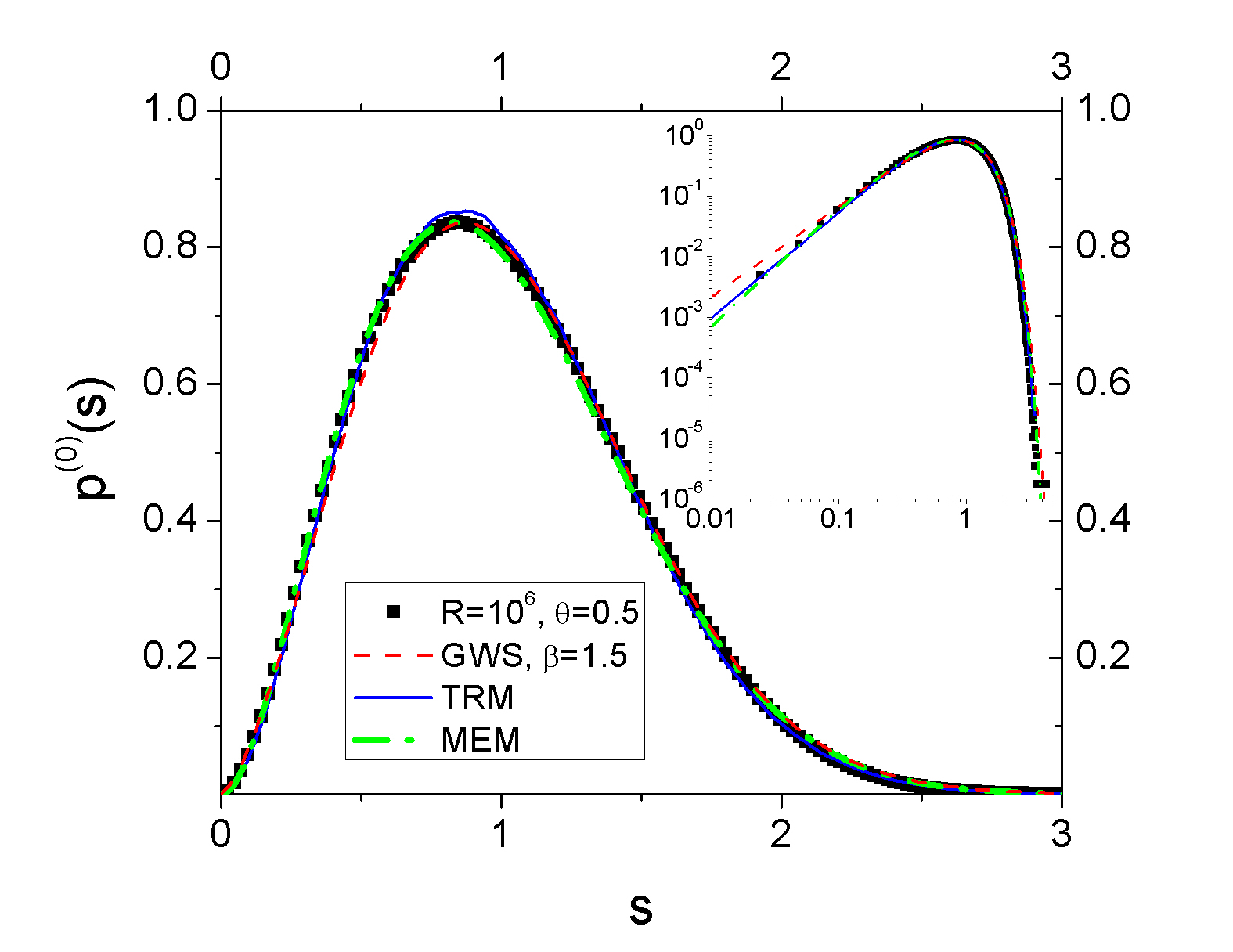}\\
(a) & (b) \\
\end{array}$
\end{center}
\caption{(Color online) The behavior of $P(s)$ and  $p^{(0)}(s)$ are shown in (a) and (b), respectively. It is clear that the GWS is an excellent approximation to these distributions for intermediate values of $s$.}
\label{gwsf}
\end{figure*}

While the main goal of this paper has been to fully characterize the spacing distribution of point islands in 1D, particularly their tails, an underlying motivation of much of our work has been to glean physical information from experimental data.  For that, the number of realizations are generally less than 1000, often significantly less.  Hence, the noise in the tails is too large to assess the effects discussed above.  Fits to data are largely determined by the range $0.5 < s < 2.5$.  In that range, the agreement between Eq.~(\ref{gws}) and our numerical results is excellent, as seen in Fig.~\ref{gwsf}. For the 1D problem with $i=1$ treated here, the fitted values of $\beta$ are 1.5 and 4 for $p^{(0)}(s)$ and $P(s)$, respectively (for details see Ref.~\cite{pimpinelli1}). As is especially clear from the inset, the fit in the tails is, unsurprisingly, not very good. According to the mean-field-like argument in Ref.~\cite{pimpinelli1}, the capture zone (CZ) distribution can be described by a Fokker-Planck equation which has $P_\beta(s)$, with $\beta=2(i+1)$ in 1D, as its stationary solution. In the associated Langevin equation
\begin{equation}
\frac{ds}{dt}=K\,\left(\frac{\beta}{s}-B\,s\right)+\eta,
\end{equation} there is a repulsive force $K\,\beta/s$ and an attractive force proportional to $s$. Here, $K$ is a kinetic coefficient and $\eta$ arises from the random component of the external pressure. The latter is responsible for the Gaussian tail of $P(s)$ while the repulsive force dominates the behavior of $P(s)$ for small values of $s$. The assumption of Gaussian decay  underestimates the attractive force which leads to a consequent underestimation of the repulsive force. For example, in the case of $p^{(0)}(s)$ this leads to $\beta=1.5$ rather than value $\beta=2$ given exactly in Eq.~(\ref{pld2}) and within error in Fig.~\ref{graph1}(b).

While the GWS does not describe well the tails of the observed distributions, it does a considerably better in the more central part of the distribution (typically $0.5<s<2$); that part is more germane to analysis of experimental data since the data in the tails is often so sparse (often below the proper 5 hits per histogram bin) that noise makes its use suspect for quantitative analysis.

We address two questions about the applicability of the GWS. First, is there a significant difference in the values of $\beta$ calculated from the whole range of data and from the reduced interval $0.5<s<2$?  We find $\beta(\rm all)=4.229$ for the entire set of data and $\beta(\rm sig)=4.222$ for the reduced range. Since this difference is insignificant, $\beta$ can be calculated from just the data in the central region. Second, is the GWS is a reasonable approximation of $P(s)$ in the range $0.5<s<2$, especially in comparison with other analytical models?  While there is no unique measure of ``goodness of fit",  we adopt as our quantitative diagnostic the standard Pearson's reduced chi-square statistic \cite{rice}:
\begin{equation}
\mathrm{\tilde{\chi}^2}=\frac{1}{\mathsf{N}- {\sf n} -1}\sum^n_{j=1}\frac{\left(P^{\rm obs}(s_j)-P^{\rm exp}(s_j)\right)^2 }{P^{\rm exp}(s_j)},
\label{chisq}
\end{equation}
where $P^{\rm obs}(s)$ is the probability of the capture zones with size $s$ observed from the numerical data, and  $P^{\rm exp}(s)$ is that expected from the analytical model; in the prefactor, distinguishing this reduced $\mathrm{\tilde{\chi}^2}$ from $\mathrm{\chi}^2$, {\sf \textit {N}} is the number of bins (typically 239) and {\sf \textit {n}} is the number of fitting parameters.  Another possible metric, e.g., would be the area between each model curve and the numerical data, i.e., the sum
over $|P^{\rm obs}(s_j)-P^{\rm exp}(s_j)|$.

In Table I we use $\mathrm{\tilde{\chi}^2}$ to compare fits obtained from four different analytical models for the capture zone distribution in 1D. We consider the GWS with $\beta=4$ suggested in Ref.~\cite{pimpinelli} (GWS-0), the best fit using the single-parameter GWS (GWS-1), the MEM+IIA result given by Eq.~(\ref{cz1D}), the single-parameter gamma distribution ($\Gamma$D)
\begin{equation}\label{gamma1}
\Pi_{\alpha}(s)=\frac{\alpha^\alpha}{\Gamma(\alpha)}s^{\alpha-1}e^{-\alpha s},
\end{equation}
the generalized gamma distribution (G$\Gamma$D)
\begin{equation}\label{gammagen}
\Pi_{\beta,\nu}(s)=A_{\beta}s^{\beta}e^{-B_{\beta}s^{\nu}},
\end{equation}
and finally our TRM. Note that with $\nu=2$ Eq.~(\ref{gammagen}) reduces to the GWS while for $\nu=1$ it becomes $\Gamma$D. Clearly, the reduction factor in transforming $\mathrm{\chi}^2$ to $\mathrm{\tilde{\chi}^2}$ in Eq.~(\ref{chisq}) is here a trivial multiplicative factor since {\sf \textit {N}} $\gg$
{\sf \textit {n}}.  However, in addition to computing $\mathrm{\tilde{\chi}^2}$ using the whole range of the data, we also find $\mathrm{\tilde{\chi}^2}$ using just the data in the interval $0.5<s<2$.  In this process, the reduction factor is significant in making subsequent comparisons.
\begin{table}[hbtp]
\caption{Values for the $\tilde{\chi}^2(\rm all)$ and $\tilde{\chi}^2(\rm sig)$ for four different analytical models in 1D. In the case of the $\Gamma$D we have used $\beta=\alpha-1$; thus, $\alpha
\approx 2 \beta +1$ of GWS-1, as noted in Ref.~\cite{rajesh}.}
\begin{center}
    \begin{tabular}{ | c | c | c | c | c | c | c | c | c |}
    \hline
      & $\Gamma$D   & GWS-0     & GWS-1       & G$\Gamma$D   & MEM+IIA    & TRM       \\ \hline
     $10^{3}\, \tilde{\chi}^2(\rm all)$  &  5.155   & 0.429       & 0.145  &  0.038     & 1.691   & 0.161\\ \hline
     $10^{3}\, \tilde{\chi}^2(\rm sig)$ &   2.483  & 0.353    & 0.218  &  0.0285   & 2.076   & 0.162\\ \hline
     $\nu$          & 1      & 2             & 2           & 1.815      & NA           & NA        \\ \hline
     $\beta$        & 8.612   & 4            & 4.229       & 4.650      & NA           & NA        \\ \hline
     \hline
    \end{tabular}
\end{center}
\label{table1D}
\end{table}

Based on Eq.~(\ref{chisq}) the GWS is better than the $\Gamma$D and the MEM+IIA but worse than the  G$\Gamma$D and the TRM. Additionally, the GWS gives a value of $\beta$ similar to that of G$\Gamma$D. We recall that in the MEM+IIA model, $P(s)\sim s^5$, which is closer to the G$\Gamma$D result than the $\beta$ of GWS. With the extra adjustable parameter, the G$\Gamma$D gives the smallest $\tilde{\chi}^2$, but at the price of non-integer values for $\beta$ and $\nu$.  Not only is the TRM excellent overall, it has nearly identical $\mathrm{\tilde{\chi}^2}$ for the full and central ranges of $s$.  It is curious that for GWS-1, $\mathrm{\tilde{\chi}^2}$ is larger for the central range than for the full one.

It is noteworthy that the mean-field reasoning of Ref.~\cite{pimpinelli1} succeeds in 1D.  This mean-field argument is based on knowledge of $n(r)$, which in 1D is $n(x,y)=(2\,R)^{-1}x(y-x)$.  Similarly, Shi et al.\cite{amar3} found that in 1D $P_4(s)$ only modestly underestimated the asymptotic peak height of the distribution for their two versions of the point island model with irreversible growth. The peak of $P(s)$ obtained from the TRM is approximately 1.29, which is close to the value 1.31  reported in Ref.~\cite{amar3} for the limit $R\rightarrow\infty$.

\subsection{Extension to 2D}

\begin{figure}[t]
\begin{center}
\includegraphics[scale=0.35]{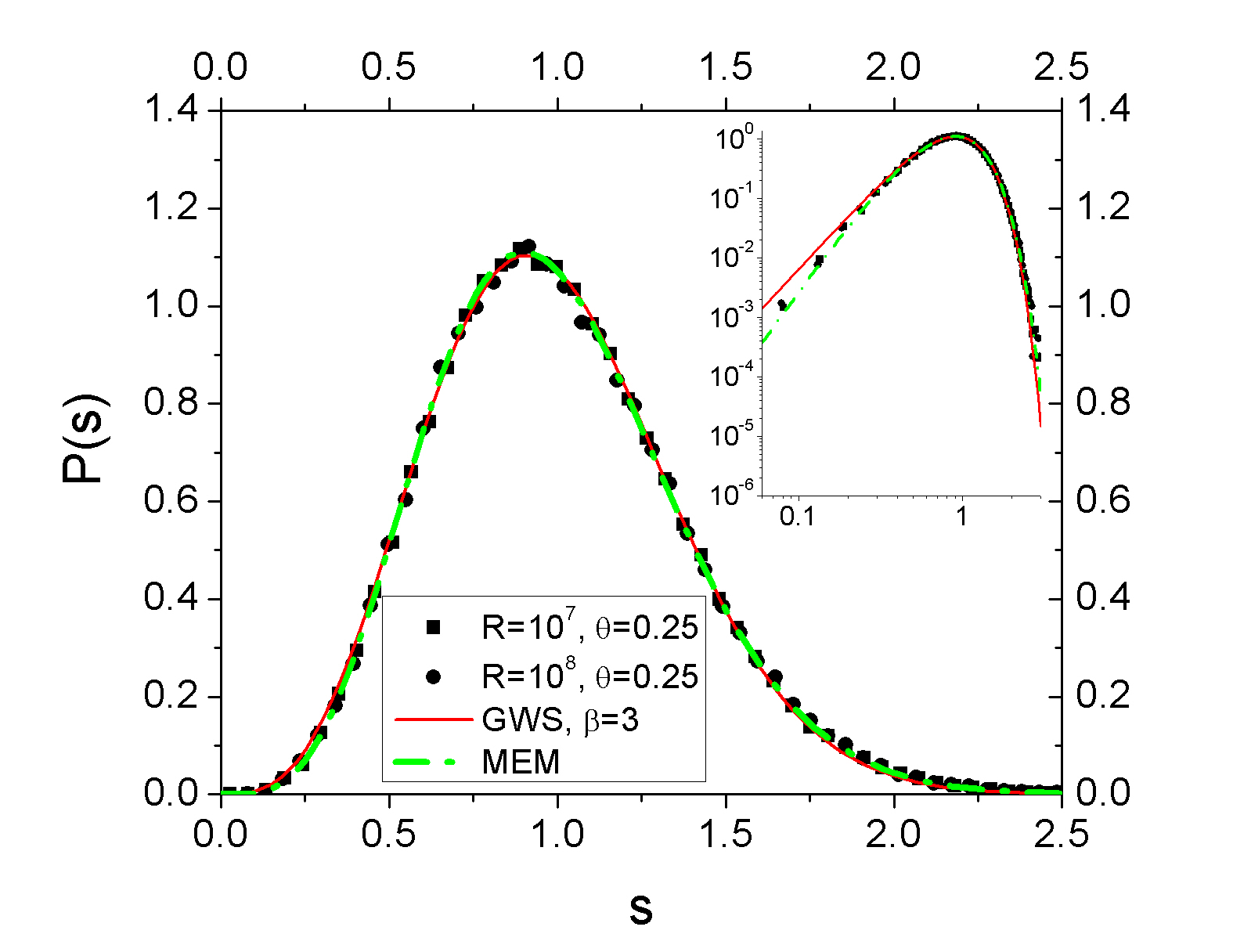}

\end{center}
\caption{(Color online) Capture zone distribution in 2D with $i=1$. The GWS describes correctly the behavior of $P(s)$ for intermediate values of $s$. The maximum entropy method gives an excellent approximation for $P(s)$ even for large and small values of $s$.}
\label{f3}
\end{figure}

The maximum entropy method can also be applied in the case of deposition in 2D. We have two normalization conditions on the capture zone distribution $P(s)$: $\int^{\infty}_0ds\,P(s)=1$ and $\int^{\infty}_0ds\,s\,P(s)=1$. We can find an additional condition: It is long known that in the aggregation regime, the densities of monomers and islands evolve as $N_1\sim\theta^{-\frac{1}{3}}$ and $N\sim\theta^{\frac{1}{3}}$, respectively \cite{lam}. On the other hand, the density of monomers,  $n(r)$, inside of a circular capture zone with radius $r_c$ is given approximately by \cite{evansgama}
\begin{equation}\label{nr}
n(r)=0.25\,R^{-1}\left(r_{isl}^2-r^2\right)+0.5\,r_c^2\,R ^{-1}\ln\left(\frac{r}{r_{isl}}\right),
\end{equation}
where $r_{isl}$ is the radius of the island and $r$ is the distance from the cell center. As before, $R=D/F$. Then, neglecting logarithmic corrections, the number of monomers inside a capture zone with radius $r_c$ is proportional to $r_c^4$, i.e., the square of the area of the capture zone. Consequently $N_1\sim \langle S^2\rangle N$. The analog of Eq.~(\ref{q3}) for this case is
\begin{equation}
\tilde{\mu}_2\sim N_1\,N,
\end{equation}
where $\tilde{\mu}_2$ is the second moment of $P(s)$, which is constant in the aggregation regime. Finally, since $P(0)=0$, it is reasonable to write
\begin{equation}\label{pmem2}
P_{\rm MEM}^{(2D)}(s)\approx A\,s^{\beta}\,e^{-B\,s^2-C\,s},
\end{equation}
where $A$, $B$ and $C$ are constants. Fig.~\ref{f3} shows a comparison between Eq.~(\ref{pmem2}) and numerical simulations of island nucleation in 2D.  According to Ref.~\cite{pimpinelli}, as well as Ref.~\cite{evans1}, $P_3(s)$, the GWS with $\beta=3 = i+2$ (rather than the original mean field result $\beta=2 = i+1$ in Ref.~\cite{pimpinelli1}) is a good approximation for $P(s)$; hence, we include it in Fig.~\ref{f3}. In 2D the GWS evidently approximates $P(s)$ well for intermediate values of $s$ but differs substantially from the numerical simulations for small and large values of $s$. However, Eq.~(\ref{pmem2}) gives an even better approximation for $P(s)$ over the whole range of $s$ (see Fig.~\ref{f3}). Based on the numerical results of Ref.~\cite{evans1}, we use $\beta=4$  to calculate the $P_{\rm MEM}^{(2D)}(s)$ shown in Fig.~\ref{f3}.

For the GWS, we compare again the values of $\beta$ calculated from the fits using the whole range of data and the restricted range $0.5<s<2$: $\beta(\rm all)\approx3.07$ and $\beta(\rm sig)\approx3.04$. Again, we evidently can calculate $\beta$ just as well using only the data from the restricted range. In Table II we quantify with the reduced Pearson's $\mathrm{\tilde{\chi}^2}$ the goodness of the fits obtained from some analytical models for the capture zone distribution in 2D. In particular, we consider the GWS with $\beta=3$ suggested in Ref.~\cite{pimpinelli}, Eq.~(\ref{pmem2}) given by the MEM, and the generalized gamma distribution (G$\Gamma$D). For the G$\Gamma$D, we consider not only the best fit (yielding  $\beta\approx3.86$ and $\nu\approx1.59$) but also the particular parametrization $\beta=4$ and $\nu=1.5$ suggested by the Evans group \cite{evans1} and hence denoted G$\Gamma$E. The one-parameter gamma distribution ($\Gamma$D) is also included. Again we compute $\chi^2(\rm all)$ using the whole range of the data and $\chi^2(\rm sig)$ using just the data in the interval $0.5<s<2$. Our results are summarized in Table II.

\begin{table}[hbtp]
\caption{Values for the $\tilde{\chi}^2(\rm all)$ and $\tilde{\chi}^2(\rm sig)$ for four different analytical models in 2D.}
\begin{center}
    \begin{tabular}{ | c | c | c | c | c | c | c |}
    \hline
      & $\Gamma$D      &GWS-0    & GWS-1        & G$\Gamma$E     &     G$\Gamma$D   & MEM \\ \hline
     $10^{3}\, \tilde{\chi}^2(\rm all)$    &    3.010     & 1.660 & 1.726      &  0.402      &   0.334     & 0.518\\ \hline
     $10^{3}\, \tilde{\chi}^2(\rm sig)$       & 1.722    &0.826 &  0.873       &  0.381            & 0.287     & 0.294\\ \hline
     $\nu$                                      &   1        & 2        & 2             & 1.5                     & 1.585        & NA \\ \hline
     $\beta$                                    &  6.277        & 3        & 3.065       & 4                   & 3.860          & 4 \\ \hline
     \hline
    \end{tabular}
\end{center}
\label{table2D}
\end{table}

Clearly the GWS is inadequate outside the range $0.5<s<2$, the $\mathrm{\tilde{\chi}^2}$(all) given by the GWS is around five and four times as large as the ones given by the G$\Gamma$D and G$\Gamma$E($\nu=1.5$), respectively. In the 2D case, the values of $\mathrm{\tilde{\chi}^2}$ of the GWS are closer to the ones given by the best fit possible of G$\Gamma$D than it does in 1D. However, the discrepancies between the GWS and the numerical data yield an underestimation of the key parameter $\beta$. In fact, the difference between the accepted values of $\beta$ and the ones given by the GWS in 2D is 25\%. The $\mathrm{\tilde{\chi}^2}(\rm sig)$ for the GWS is around 3 and 2 times as large as the ones given by the G$\Gamma$D and G$\Gamma$E, respectively. This suggests that the GWS can be used as a first approximation for the CZ distribution in the range $0.5<s<2$. One of the main goals of the experiments on epitaxial growth is to determine the ``critical nucleus"  size $i$ from the experimental data. Then, in order to estimate $i$, it is necessary to assume an approximate functional form for $P(s)$ and then make a fit to find an approximate value for $i$. Unfortunately, the functional form of $P(s)$ depends on $i$ in a non trivial way. The GWS can be used as a first approximation to determine $i$ from experimental data because most of these data are inside the range where the GWS is a reasonable approximation.

We emphasize that the different analytic fitting functions must be expected to yield different values of $\beta$, i.e. the exponent associated with the power-law rise for $s \ll 1$.  In Table \ref{table2D} the values of $\beta$ for the two GWS fits is considerably smaller (by about 1) than the $\beta$s from G$\Gamma$ and MEM functions, even though all provide decent accounting of the central region.  The functional forms incorporate different underlying assumptions, and only the GWS has been related, if approximately, to the critical nucleus $i$.  Hence, there is considerable subtlety to extracting $i$ from the values of $\beta$ obtained by fits to expressions other than GWS.

In 2D $\tilde{\chi}^2(\rm sig)$ is invariably smaller than $\tilde{\chi}^2(\rm all)$.  However, for G$\Gamma$D and G$\Gamma$E, they are relatively close, indicating the most consistent fit over the whole range, while the GWS is significantly better in the central region.  The GWS is notably better than the $\Gamma$D, often used to analyze froths \cite{Weaire} and quantum dot distributions \cite{dot}.

In contrast to 1D, the mean-field reasoning of Ref.~\cite{pimpinelli1}, with nucleation probability $\propto n^2(r)$ (and constant $n(r)$), is not viable in 2D.  In 2D the better estimate of $n(r)$ given in Eq.~(\ref{nr}) comes from solving the diffusion equation; however the radial extent of a CZ fluctuates significantly (as a function of polar angle) around its mean value $r_c=(S/ \pi)^{1/2}$. Furthermore, the island is not usually at the geometric center of the CZ. Hence, Eq.~(\ref{nr}) generally does not adequately approximate the local density of monomers inside a CZ.  In 1D newly nucleated CZs obtain their ``area" from just two existing CZs, while in 2D several existing CZs often contribute.  Most of these nucleations occur near the CZ boundaries, which generally are not well described as circles.

\section{Conclusions}

\begin{figure*}[t]
\begin{center}
$\begin{array}{cc}
\includegraphics[scale=0.45]{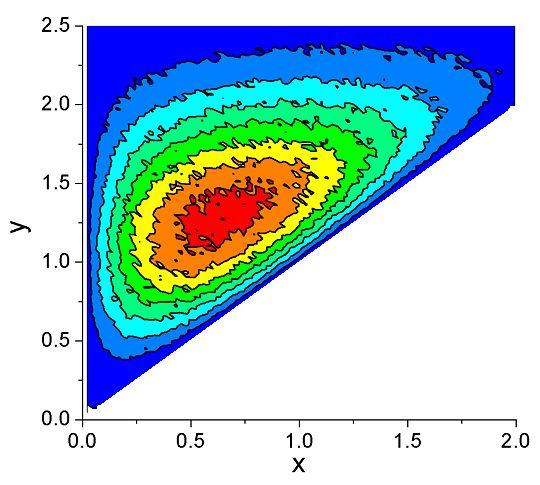}&
\includegraphics[scale=0.45]{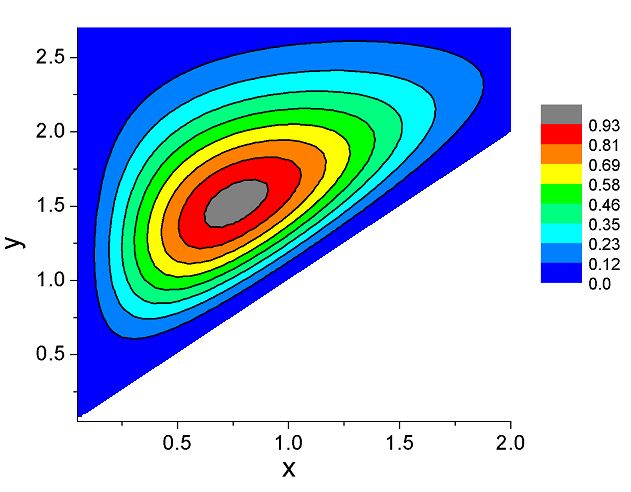}\\
(a) & (b) \\
\end{array}$
\end{center}
\caption{(Color online) Behavior of $p^{XY}_n(x,y)$ from (a) the numerical simulation and (b) Eq.~(\ref{epnxy}) with $\gamma=4$. The data used in this figure were generated with $R=10^6$ and $\theta=0.5$}
\label{pncxy}
\end{figure*}

BM \cite{blackman} proposed that $p^{Y}_{n}(y)/F_M(y)\propto y^{\gamma}$ with $\gamma=5$. However, we find that $\gamma=3$ and 4 give much better agreement with our numerical simulation. For example, if we choose $\gamma=4$ from Eqs.~(\ref{epnxy}) and (\ref{paprox}) it is possible to write an analytical expression for the probability density $p^{XY}_n(x,y)$ in terms of $p^{(0)}(s)$. In Fig.~\ref{pncxy} we show the results of calculating $p^{XY}_n(x,y)$ numerically compared with our analytical model. The contour plots are quite similar. However, there is a small difference in the location $(x,y)$ of the maximum value of $p^{XY}_n(x,y)$: In the analytical model it is located at approximately (0.7,1.5), while in the numerical simulation it is at (0.7,1.3).

However, neither $\gamma=3$ nor $\gamma=4$ describe the statistical behavior of the system for large values of $s$. Furthermore, the TRM gives an excellent quantitative description of the statistical behavior of the point-island model even for large values of $s$, as shown in Fig.~\ref{p01D}. This  means that a complete description of the system requires to take into account that $\gamma$ is a function of $s$, i.e., the probability to choose a gap $p^Y_n(y)$ is different for large and small gaps, see Fig.~\ref{pysmall}.

We find that the spacing distribution $p^{(0)}(s)$ decays like $\exp(-B\,s^3)$ instead of $\exp(-B\,s^5)$ as in BM or $\exp(-B\,s^2)$ as in Ref.~\cite{pimpinelli1}. It seems that this is consequence of the confinement of the diffusive monomers within the gaps. An analysis of Eq.~(\ref{difp}) shows that the tail depends exclusively on the probability to choose the gap $p^Y_n(y)$. This suggests that the statistical behavior of the system for large values of $s$ is dominated by the breakup of the biggest gaps where $\gamma\approx3$. Meanwhile, the behavior of $p^{(0)}(s)$ for small values of $s$ depends on the probability to choose the position of the nucleation inside the gap $p^{\Lambda}_n(\lambda)$.

Fig.~\ref{ps} shows that the maximum entropy principle can be used to find an excellent analytical approximation for $p^{(0)}(s)$. Equation (\ref{paprox}) gives an excellent fit of the numerical data, even for large and small values of $s$, in fact better than other approximate expressions \cite{blackman,pimpinelli1}. We find that by using the IIA and Eq.~(\ref{paprox}), we can  easily  handle expressions for $p^{(k)}(s)$ with $k>0$. Those expressions accurately describe the numerical data, even the weak oscillation in $G(r)$. Apparently, for small values of $s$ the spacing distribution function satisfies $p^{(k)}(s)\propto s^{3\,k+2}$. We show that Eq.~(\ref{ed1d}) and (\ref{paprox}) provide an excellent approximation for $p^{X}_n(s)$, as seen in Fig.~\ref{ps}. While $p^{\Lambda}_n(\lambda)\propto n^{2}(\lambda)$ leads to $G(r)\propto g^{2}(r)$ for small values of $r$, this relation is not satisfied for values of $r$ near the maximum of $G(r)$.

From our previous results, it is possible to calculate the capture number, $\sigma_s$, of an island with size $s$ in the aggregation regime. As usual \cite{blackman,amar6}, $\sigma_s$ can be determined from
\begin{equation}
\sigma_s=\frac{1}{N_1}\left(\left.\frac{dn_1(x,y)}{dx}\right|_{x=0^+} - \left.\frac{dn_1(x,y)}{dx}\right|_{x=0^-}\right),
\end{equation}
where $N_1$ is the density of monomers and $\left.(dn_1(x,y)/dx)\right|_{x=0}\!{\scriptscriptstyle +[-]}$ represents the derivative of $n_1(x,y)$ at the right [left] boundary of the island. By using the approximation given previously for $n_1(x,y)$, we found
\begin{equation}
\sigma_s=\frac{y^{CZ}}{N_1\,R},
\end{equation}
where $y^{CZ}$ is the length of the capture zone. In the point-island model the capture numbers naturally do not depend on the size of the island. This result also implies that the distribution of $\sigma_s$ has the same form of $p^{(1)}(s)$ and not that of $p^{(0)}(s)$ as was suggest in Ref.~\cite{amar5}. There the authors supposed that the size of the gap at the left and the right of an island are equal. A better first approximation for the distribution of $\sigma_s$ can be obtained if we calculate $p^{(1)}(s)$ from $p^{(0)}(s)$ through a convolution product, as in Fig.~\ref{ps}(a).

A major result of our analysis, consistent with Refs.~\cite{evans1,pimpinelli1}, is that obtaining an appropriate description of the nucleation mechanism is the crucial ingredient to arriving at an excellent approximation for the spacing distribution functions of the point-island model.
Finally, we emphasize that in spite of its mathematical simplicity, the GWS with the suitable selection of $\beta$ is a good approximation for $P(s)$ in 1D and 2D. Because of this, it is reasonable to use the GWS to analyze experimental results for epitaxial growth.

\section*{Acknowledgments}
We thank Prof. E. Slud for enlightening comments about ``goodness of fit."  This work was supported by the NSF-MRSEC at the University of Maryland, Grant No.\ DMR 05-20471 and a DOE CMCSN team grant, with ancillary support from the Center for Nanophysics and Advanced Materials (CNAM).

\appendix
\section{Behavior of $p^{(0)}(s)$ for small and large values of $s$}\label{app1}
An expansion of Eq.~(\ref{pn1}) around $s=0$ shows that $p^{X}_n(s)\approx (30/\mu_{\gamma})s^2$ for small values of $s$. By using this result in Eq.~(\ref{difp}), we found  $p^{(0)}(s)\approx (15/\mu_{\gamma})s^2$. This result is a consequence of the fact that $p^{\Lambda}(\lambda)\approx 30\,\lambda^2$ as $\lambda\rightarrow0$.

From our numerical results, it is reasonable to propose $p^{(0)}(s)\approx A\,s^{\alpha}\,e^{-B\,s^{\rho}}$ for large values of $s$, where $A$, $B$, $\alpha$ and $\rho$ are constants. By using this ansatz in Eq.~(\ref{pn1}) it is clear that for large values of $s$, $p^{X}_n(s) \ll p^{(0)}(s)$. Thus, Eq.~(\ref{difp}) takes the form
\begin{equation}
(2+\alpha)-B\,\rho\,s^{\rho}+\frac{s^{\gamma}}{\mu_{\gamma}}\approx0,
\end{equation}
which implies $\alpha=-2$, $B=(\rho\,\mu_{\gamma})^{-1}$ and $\rho=\gamma$. The behavior of $p^{(0)}(s)$ for large values of $s$ is fully determined by the probability to choose the gap to fragment.



\begin{thebibliography}{}
\addcontentsline{toc}{chapter}{References}
\bibitem{blackman} J. A. Blackman and P. A. Mulheran, Phys. Rev. B. \textbf{54}, 11681 (1996).
\bibitem{amar} J. G. Amar and F. Family, Phys. Rev. B \textbf{50}, 8781 (1994).
\bibitem{mulheran} P. A. Mulheran and D. A. Robbie, Europhys. Lett. \textbf{49}, 617 (2000).
\bibitem{amar1} J. G. Amar and F. Family, Phys. Rev. Lett. \textbf{74}, 2066 (1995).
\bibitem{ratsch} C. Ratsch, A. Zangwill, P. \v{S}milauer and D. D. Vvedensky, Phys. Rev. Lett. \textbf{72}, 3194 (1994).
\bibitem{amar2} M. N. Popescu, J. G. Amar, and F. Family, Phys. Rev. B \textbf{64}, 205404 (2001).
\bibitem{blackman2} P. A. Mulheran and J. A. Blackman, Phys. Rev. B \textbf{53}, 10261 (1996).
\bibitem{amar3} F. Shi, Y. Shim and J. G. Amar, Phys. Rev. E \textbf{79}, 011602 (2009).
\bibitem{evans} J. W. Evans and M. C. Bartelt, Phys. Rev. B \textbf{63}, 235408 (2001).
\bibitem{amar4} M. N. Popescu, J. G. Amar and F. Family, Phys. Rev. B \textbf{58}, 1613 (1998).
\bibitem{amar5} J. G. Amar and M. N. Popescu, Phys. Rev. B \textbf{69}, 033401 (2004).
\bibitem{amar6} J. G. Amar, M. N. Popescu F. Family, Surf. Sci. \textbf{491}, 239, (2001)
\bibitem{amar7} F. Shi, Y. Shim and J. G. Amar, Phys. Rev. E \textbf{71}, 245411 (2005); \textbf{74}, 021606 (2006).
\bibitem{ratsch1} C. Ratsch, Y. Landa and R. Vardavas, Surf. Sci. \textbf{578}, 196 (2005).
\bibitem{tokar} V. I. Tokar and H. Dreyss\'e, Phys. Rev. B \textbf{80}, 161403(R) (2009).
\bibitem{evansgama} J. W. Evans and M. C. Bartelt, Phys. Rev. B \textbf{66}, 235410 (2002).
\bibitem{pimpinelli1} A. Pimpinelli and T. L. Einstein, Phys. Rev. Lett. \textbf{99}, 226102 (2007).
\bibitem{evans1} M. Li, Y. Han, and J. W. Evans, Phys. Rev. Lett. \textbf{104}, 149601 (2010).
\bibitem{pimpinelli} A. Pimpinelli and T. L. Einstein, Phys. Rev. Lett. \textbf{104}, 149602 (2010).
\bibitem{castellano} C. Castellano and P. Politi, Phys. Rev. Lett., \textbf{87}, 5 (2001).
\bibitem{vardavas} R. Vardavas, Fluctuation and Scaling in 1D Irreversible Film Growth, Ph.D Thesis (Imperial College, London, 2002), unpublished.
\bibitem{jaynes} E. T. Jaynes, Phys. Rev. \textbf{106}, 620 (1957);  \textbf{108}, 171 (1957).
\bibitem{wu} N. Wu, The Maximum Entropy Method (Springer, New York, 1997).
\bibitem{kumar} V. S. Kumar and V. Kumaran, J. Chem. Phys. \textbf{123}, 114501 (2005).

\bibitem{gonzalez4} D. L. Gonz\'alez, A. Pimpinelli and T. L. Einstein. Mean-Field Approximation for Spacing Distribution Functions in Classical Systems, preprint.



\bibitem{bales} G. S. Bales and D. C. Chrzan, Phys. Rev. B \textbf{50}, 9 (1994).
\bibitem{evansg} M. C. Bartelt and J. W. Evans,  Surf. Sci. 298, 421 (1993).
\bibitem{evansr} J. W. Evans, P. A. Thiel, and M. C. Bartelt, Surf. Sci. Rep. \textbf{61}, 12 (2006).
\bibitem{bressler} V. Bressler-Hill, S. Varma,  A. Lorke, B. Z. Nosho, P.M. Petroff and W. H. Weinberg, Phys. Rev. Lett. \textbf{74}, 16 (1995).
\bibitem{bartel} M. C. Bartelt and J. W. Evans, Phys. Rev. B \textbf{46}, 19 (1992).
\bibitem{rice} John A. Rice, Mathematical Statistics and Data Analysis, 2nd Ed. (Duxbury Press, Belmont, CA, 1995).
\bibitem{rajesh} R. Sathiyanarayanan, A.B.H. Hamouda, A. Pimpinelli, and T.L. Einstein, Phys. Rev. B \textbf{83}, 035424 (2011).
\bibitem{lam} J. G. Amar, F. Family, and P-M. Lam, Phys. Rev. B \textbf{50}, 8781 (1994).
\bibitem{Weaire} D. Weaire, J.P. Kermode, and J. Wejchert,  Philos. Mag. B \textbf{53}, L101 (1986).
\bibitem{dot} E.g., M. Fanfoni, E. Placidi, F. Arciprete, E. Orsini, F. Patella, and
A. Balzarotti, Phys. Rev. B \textbf{75}, 245312 (2007).

\end{thebibliography}
\end{document}